\documentclass{PoS}
\usepackage{subcaption}
\title{Status and performance results from NectarCAM  -- a camera for CTA medium sized telescopes}

\ShortTitle{Status and performance results from NectarCAM}

\author{\speaker{Thomas Tavernier}\\
        CEA-IRFU, Gif-Sur-Yvette, France\\
        E-mail: \email{thomas.tavernier@cea.fr}
        }
\author{Jean-Fran\c cois Glicenstein\\
        CEA-IRFU, Gif-Sur-Yvette, France\\
        E-mail: \email{jean-francois.glicenstein@cea.fr}}
\author{Fran\c cois Brun\\
        CEA-IRFU, Gif-Sur-Yvette, France\\
        E-mail: \email{francois.brun@cea.fr}}
\author{for the CTA Consortium\footnote{Full consortium list at http://cta-observatory.org}}       

\abstract{The Cherenkov Telescope Array (CTA) will be the first ground-based observatory for gamma-ray astronomy. 
With more than a hundred of 4th generation of Imaging Atmospheric Cherenkov Telescopes (IACTs) distributed in two large arrays, CTA will reach unprecedented sensitivity, angular resolution, and spectral coverage.
Three classes of IACTs -- 40 Medium-Sized Telescopes (MSTs), 8 Large-Sized Telescopes (LSTs) and 70 Small-Sized Telescopes (SSTs) -- are required to cover the full CTA energy range (20 GeV to 300 TeV).

NectarCAM is a  Cherenkov camera which is designed to equip medium sized telescopes of CTA, covering the central energy range from 100
GeV to 30 TeV, with a  field of view of 8 degrees.
It is based on a modular design with data channels using the NECTAr chip, which is equipped with both GHz
sampling Switched Capacitor Array and 12-bit Analog to Digital Converter (ADC).
The camera will comprise 265 modules, each consisting of 7 photomultiplier Tubes (PMTs) and a Front-End Board performing the data capture, 
 sending the data over the Ethernet after the trigger decision at rates up to 10 kHz.
 
This contribution provides an overview of the status of the first NectarCAM camera currently under integration in CEA Paris-Saclay (France). Furthermore, we will discuss the calibration strategies and present performance results from the CEA Paris-Saclay test bench and from the first data taken under a real sky on the prototype of medium sized telescope (MST) structure in Adlershof (Germany).}

\FullConference{36th International Cosmic Ray Conference -ICRC2019-\\
		July 24th - August 1st, 2019\\
		Madison, WI, U.S.A.}

\begin{document}

\section{Introduction}

The NectarCAM camera\cite{bib:NectarCAM2017} has a modular design with a basic element ("module") consisting of 7 photomultipliers (PMTs) associated with their readout and trigger electronics.
Each module contains a focal plane module (FPM), a front-end board (FEB) and a back-plane. The FPM is composed by seven R12992-100-05 
Hamamatsu PMTs\footnote{www.hamamatsu.com} associated high voltage and pre-amplification boards (HVPA). The PMTs are equipped with Winston cones. The front-end board includes a second amplification level, NECTAr chips which perform the readout together with the analogue to digital conversion, and the local trigger electronics (L0).
Each FEB is plugged to a back-plane board which produces the camera trigger decision (L1) processing the L0 signals of a 37 pixels region (including L0 signal from 6 neighbouring FEBs). This multi-level digital trigger system is described in detail in \cite{bib:trigger}. The L1 signal is then sent to the trigger interface board (TIB) \cite{bib:tib}, which sends back a L1-Accept (L1A) signal to all the modules if the trigger is accepted. The events are time-stamped on the TiCkS module \cite{bib:ticks}, a dedicated White Rabbit based board.  
When the L1A signal is received by the FEBs, the data are read-out in the NECTAr chips, digitized and transferred over Ethernet to a camera server where the event is build.  

The camera is currently under integration at the integration facility in CEA Paris-Saclay (France).
The full camera will be equipped with 265 modules (1855 pixels).
This paper presents calibration strategies and  performance results obtained with the partially equipped camera (61 modules) both at CEA Paris-Saclay test bench and on a prototype of Medium-Sized telescope (MST) structure at Adlershof (Germany).
The latter includes  real data taken during night time in May/June 2019.

\subsection{Dark room test bench}

The NectarCAM test bench consists of a $\sim$12 m long dark room at the NectarCAM integration hall at CEA Paris-Saclay. The camera is equipped  with the full readout, data acquisition and trigger electronics as well as its mechanical structure and cooling system. The two light sources are a Light Emitting Diode (LED) pulser and a continuous Night Sky Background (NSB) emulating  light controlled by an 
OPC Unified Architecture (OPCUA) server in the control room. The flatness of both light sources was evaluated for the full camera scale with the help of an x,y table.

\subsection{Adlershof campaign}

\begin{figure}[!h]
\centering
\includegraphics[clip,width=0.8\linewidth]{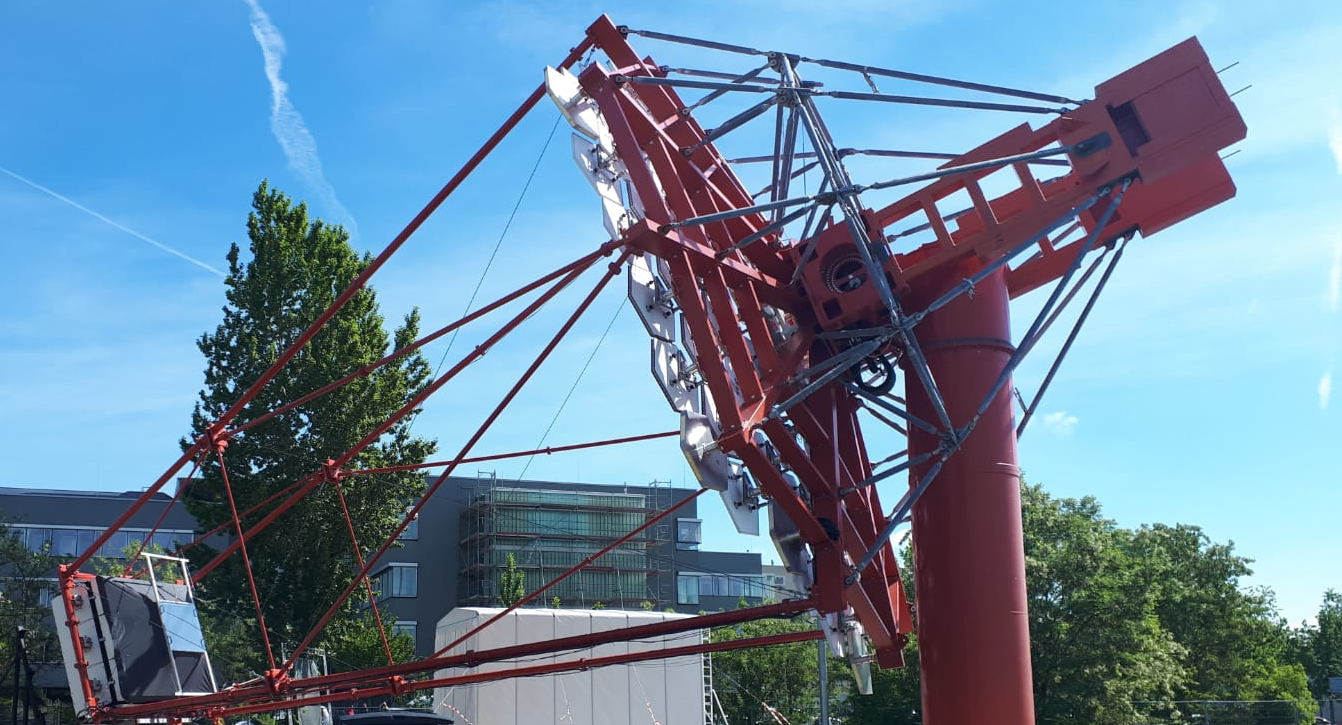}
\caption{The NectarCAM camera in the MST structure at Berlin-Adlershof.}
\label{cam}
\end{figure}

The partially equipped NectarCAM was installed at the focal plane of the MST prototype structure in Adlershof, in the  suburbs of Berlin (see fig \ref{cam}). The goal of the campaign was to perform integration tests and sky observations under realistic operating conditions. The camera was  tested during one month in May and June 2019. The light pollution (public lights) and night sky background at the Adlershof site is not optimal for astronomical observations. In order to reduce this pollution, protection devices were therefore installed, thus reducing the noise in front of the camera (filters). Adopting this setup and despite unfavourable observing conditions, successful acquisition of air shower events and calibration runs were conducted.

\section{Gain estimation and calibration}


The gain of a PMT is defined as the number of Analog to Digital Converter (ADC) counts per photo-electron (p.e.). It is evaluated from calibration runs using a low ($\lesssim$ 1 p.e.) intensity pulsed LED.
The multi photo-electron spectrum represents the ADC count distribution obtained with such runs. It's well described by a Gaussian-Poisson model where both pedestals and single p.e. components follow a Gaussian distribution. The probability density function (PDF) of such model is given by : 

\begin{equation}
p_x(x | \lambda,g,\sigma^2_{ses},\sigma^2_{ped},x_0) =
\sum^\infty_{n=0}\frac{\lambda^n e^{-\lambda}}{n!} . 
\frac{e^{-\frac{(x-gn-x_0)^2}{2(n\sigma^2_{ses}+\sigma^2_{ped})}}}
{\sqrt{2\pi(n\sigma^2_{ses}+\sigma^2_{ped})}}
\label{eq:spe}
\end{equation}

where $\lambda$ is the light intensity, $g$ the PMT's gain, $\sigma^2_{ses}$ the variance of the single photo-electron spectrum,  $\sigma^2_{ped}$ the variance of the pedestal distribution , $x_0$ the mean pedestal value.

\begin{figure}[!h]
\centering
\includegraphics[clip,width=0.5\linewidth]{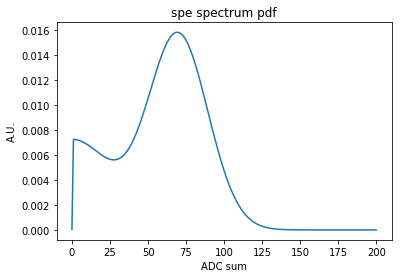}
\caption{Single p.e. spectrum using the double constrained Gaussian model. This shape is the one used for the gain estimation of the pixels.}
\label{spe2g}
\end{figure}

However, measurements at a higher gain show limitations of the Gaussian model: charge distributions exhibit a plateau between the pedestal and the single p.e. peak. This plateau can be explained by the presence of a low charge component in the distribution of the single p.e. PDF which can be described by a half Gaussian centred on 0. This single p.e. shape is shown in Fig. \ref{spe2g}.

For a given PDF of the single p.e. (noted spe($x$)), the multi photo-electron spectrum becomes :

\begin{equation}
p_x(x| \lambda,\sigma^2_{ped},x_0, \mathrm{spe}(x)) =
\sum^\infty_{n=0}\frac{\lambda^n e^{-\lambda}}{n!} . 
\frac{e^{-\frac{(x-x_0)^2}{2\sigma^2_{ped}}}}
{\sqrt{2\pi\sigma^2_{ped}}}  
\underbrace{ \circledast \mathrm{spe} \circledast \cdots \circledast \mathrm{spe}}_{n \times} ~ (x)
\label{eq:spe}
\end{equation}
The gain is then defined as the mean of the single photo-electron distribution $g ~ = ~ <\mathrm{spe}(x)>$.

At the nominal high voltage, there is no statistical evidence to discriminate between the two hypothesis. Studies performed at larger values of high voltage show that the two-component model is stable with the high voltage used and can be extrapolated to the nominal gain of the NectarCAM PMTs. The multi photo-electron spectrum is then fitted using a double constrained Gaussian-Poisson model where the ratio of the standard deviation and sum between the two Gaussians is fixed to the value found at higher gain.

\begin{figure}[!h]
\centering
\begin{subfigure}{.49\textwidth}
\centering
\includegraphics[clip,width=1.\linewidth]{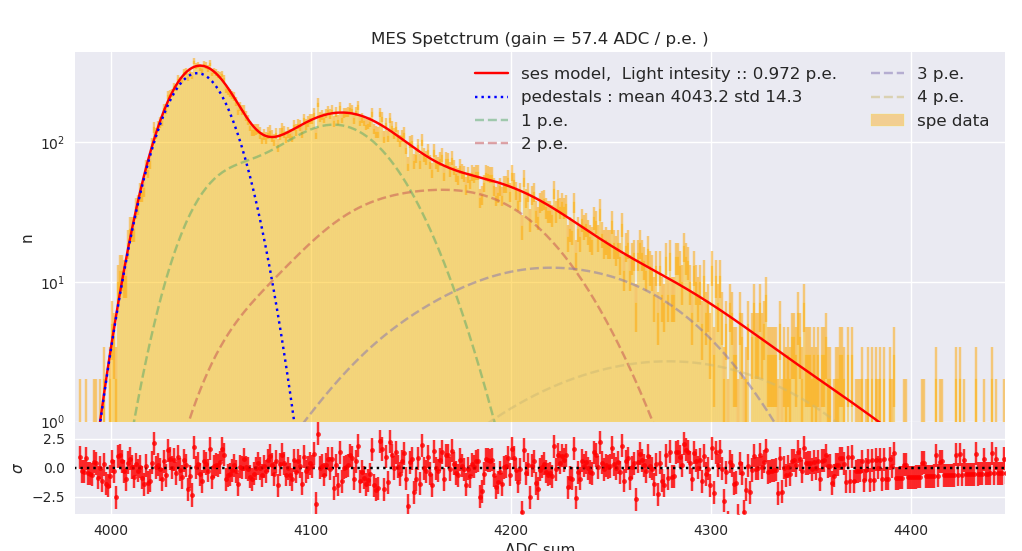}
\caption{}
\label{1gmes}
\end{subfigure}\hfill%
\begin{subfigure}{.49\textwidth}
  \centering
  \includegraphics[clip,width=.9\linewidth]{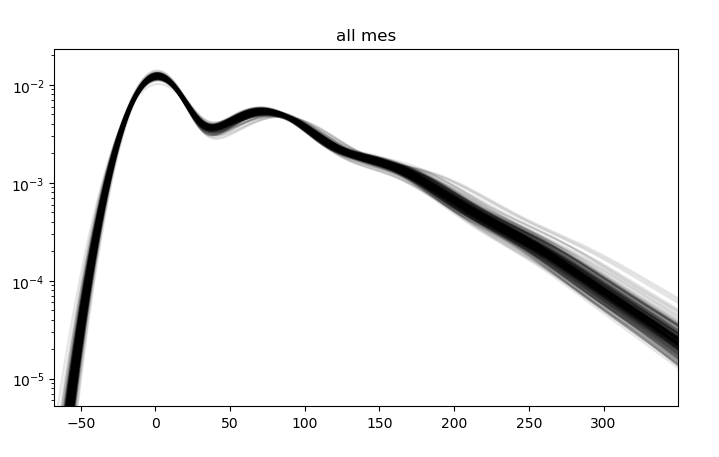}
  \caption{}
  \label{2gmes}
\end{subfigure}
\caption{Pixel multi-electron spectrum. The charge is obtained by using an integration window of 16 time slots (-6 ns +10 ns around the mean pulse position). The distribution is fitted using double constrained Gaussian-Poisson model. Figure (a) shows an example for one pixel where component from 1, 2, 3 and 4 p.e. are explicitly shown. Figure (b) Shows the superposition of the fitted multi-electron spectrum for all the pixels.}
\label{mess}
\end{figure}

\begin{figure}[!h]
\centering
\begin{subfigure}{.49\textwidth}
\centering
\includegraphics[clip,width=1.1\linewidth]{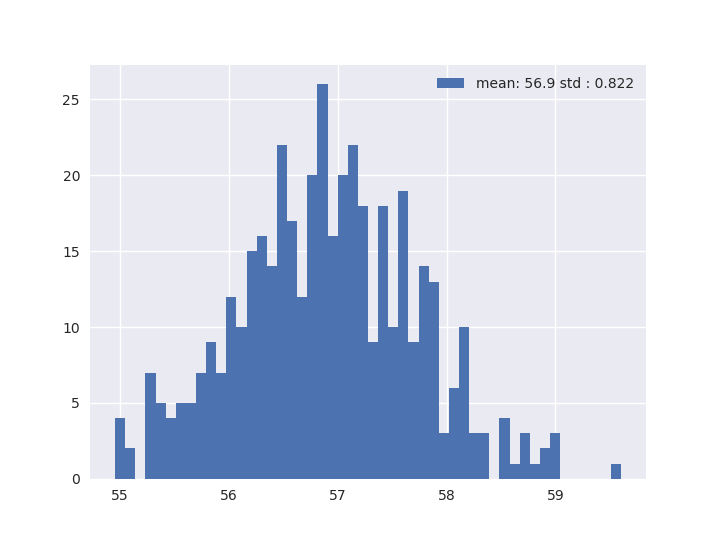}
\caption{}
\label{1gmes}
\end{subfigure}\hfill%
\begin{subfigure}{.49\textwidth}
  \centering
  \includegraphics[clip,width=1.1\linewidth]{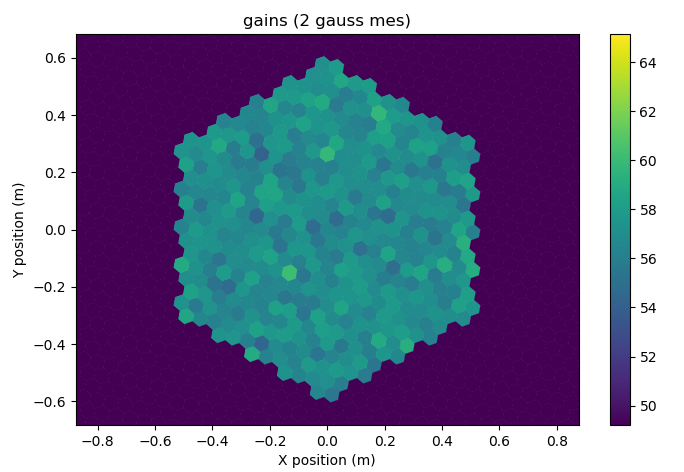}
  \caption{}
  \label{2gmes}
\end{subfigure}
\caption{Gain distributions of the camera PMT's. figure (a) shows the 1D distribution and figure (b) the gain value in the camera frame. Relative gains dispersion is $\frac{\sigma_g}{g_\mathrm{mean}} \sim 0.015$.
}
\label{ghvdist}
\end{figure}

The calibration of the PMT's gains is done by using measurements at different high voltage values. These value are then used to derive a model of the gain as a function of the high voltage for each pixel. Figure \ref{ghvdist} shows the gain distributions after calibration. The relative gain dispersion obtained with the calibration is $ \sim 1.5 $ \%. 

\section{Timing estimation and calibration}

The estimation on the light's arrival time in all the pixels of a single camera image provides an additional and valuable information. This information may significantly help to reduce the noise in shower images and may improve the image cleaning, the reconstruction of the primary particle parameters or the gamma vs background discrimination. In this section we discuss the timing accuracy and systematic uncertainties in the NectarCAM camera. The time of maximum (TOM) of the Cherenkov pulse is estimated  via a maximum likelihood method. 


\subsection{TOM evaluation Method}
To evaluate the TOM of the pulse we first define a pulse shape function $f_{\mathrm{s}}(t)$
which is extracted from real data with a charge of few tens of photo-electrons and normalised to the value of a single p.e.. The time shift $\Delta t$ of the TOMs between the model and the data is then estimated through the maximization of the log likelihood function : 
\begin{equation}
\mathrm{ln}(\mathcal{L}_{\Delta t}) =  \sum_{i=1}^{N_\mathrm{sample}} n_i . \mathrm{ln}\left(\int_{t_i}^{t_i+1} Q . f_{\mathrm{s}}(t + {\Delta t})+ \mathrm{Ped} \quad   dt \right)
\label{TOMlike}
\end{equation}

where Q is the charge extracted from the waveform, Ped is the pedestal value and $n_i$ represent the data samples of the waveform.

\subsection{Trigger accept calibration}

\begin{figure}[!h]
\centering
\begin{subfigure}{.49\textwidth}
\centering
\includegraphics[clip,width=1.1\linewidth]{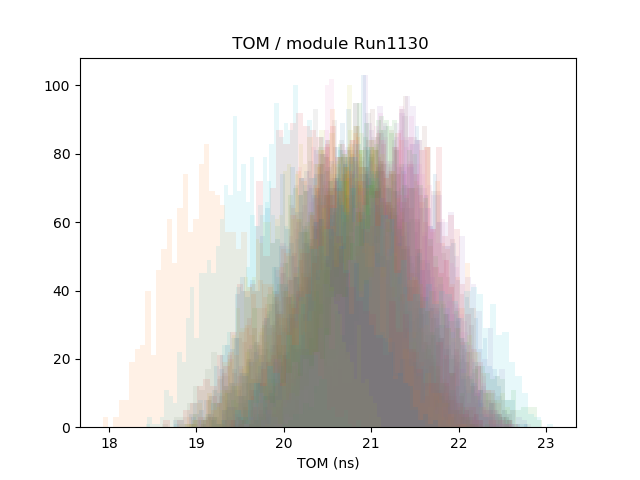}
\caption{before calibration}
\label{1gmes}
\end{subfigure}\hfill%
\begin{subfigure}{.49\textwidth}
  \centering
  \includegraphics[clip,width=1.1\linewidth]{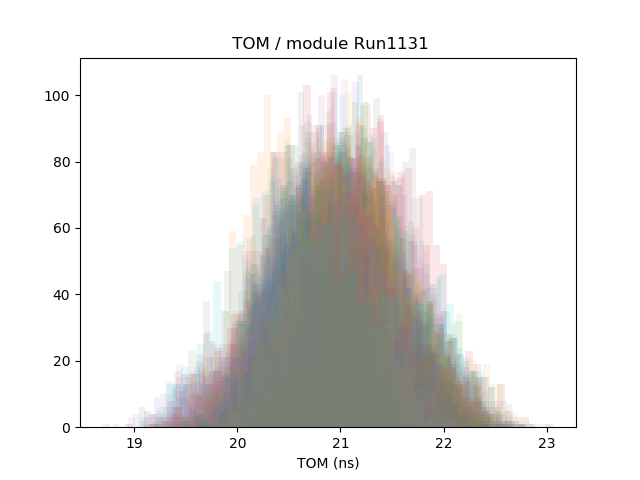}
  \caption{after calibration}
  \label{2gmes}
\end{subfigure}
\caption{
Distribution of the mean TOM in the camera before and after calibration. Each module is represented by a different color. 
}
\label{TOMCam}
\end{figure}

One of the sources of the timing offsets between pixels is the time dispersion (for each module) during the travel of L1A trigger signal between the TIB and the modules. An adjustable delay in the readout system is provided for this purpose. The mean time of maximum for each pixel is evaluated using a pulsed LED light source at $\sim 80 p.e.$ which also sends a trigger signal to the TIB. Figure \ref{TOMCam} shows the mean TOM distribution before and after calibration. After the calibration, the standard deviation of the mean TOM distribution in the camera is $\sigma_\mathrm{TOM} \sim 0.43$ ns.

\begin{figure}[!h]
\centering
\includegraphics[clip,width=0.5\linewidth]{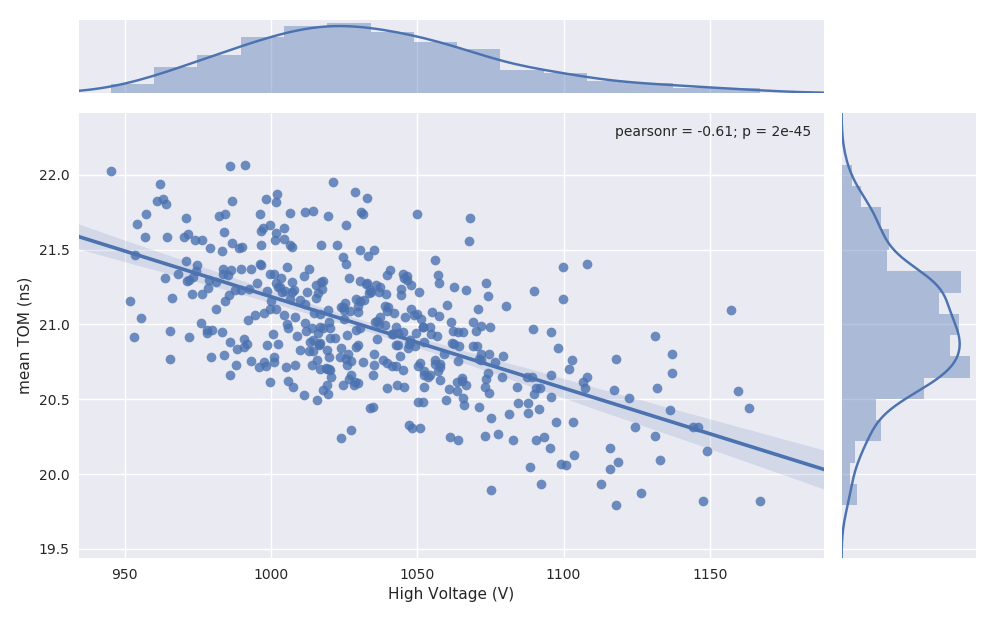}
\caption{2D distribution of the mean time of arrival in each pixel as a function the PMT high voltage.Top and right panels respectively show the HV and TOM distribution.}
\label{TOMvshv}
\end{figure}

After the trigger accept calibration, another source of timing offsets between pixels is the HV dependence of the time transfer of the electron burst in the PMT. This dependence is shown in figure \ref{TOMvshv} and can be corrected at the analysis level to reach a standard deviation of the mean TOM distribution in the camera of $\sigma_\mathrm{TOM} \sim 0.34$ ns.

\subsection{Timing measurements in shower data}

\begin{figure}[!h]
\centering
\includegraphics[clip,width=1.0\linewidth]{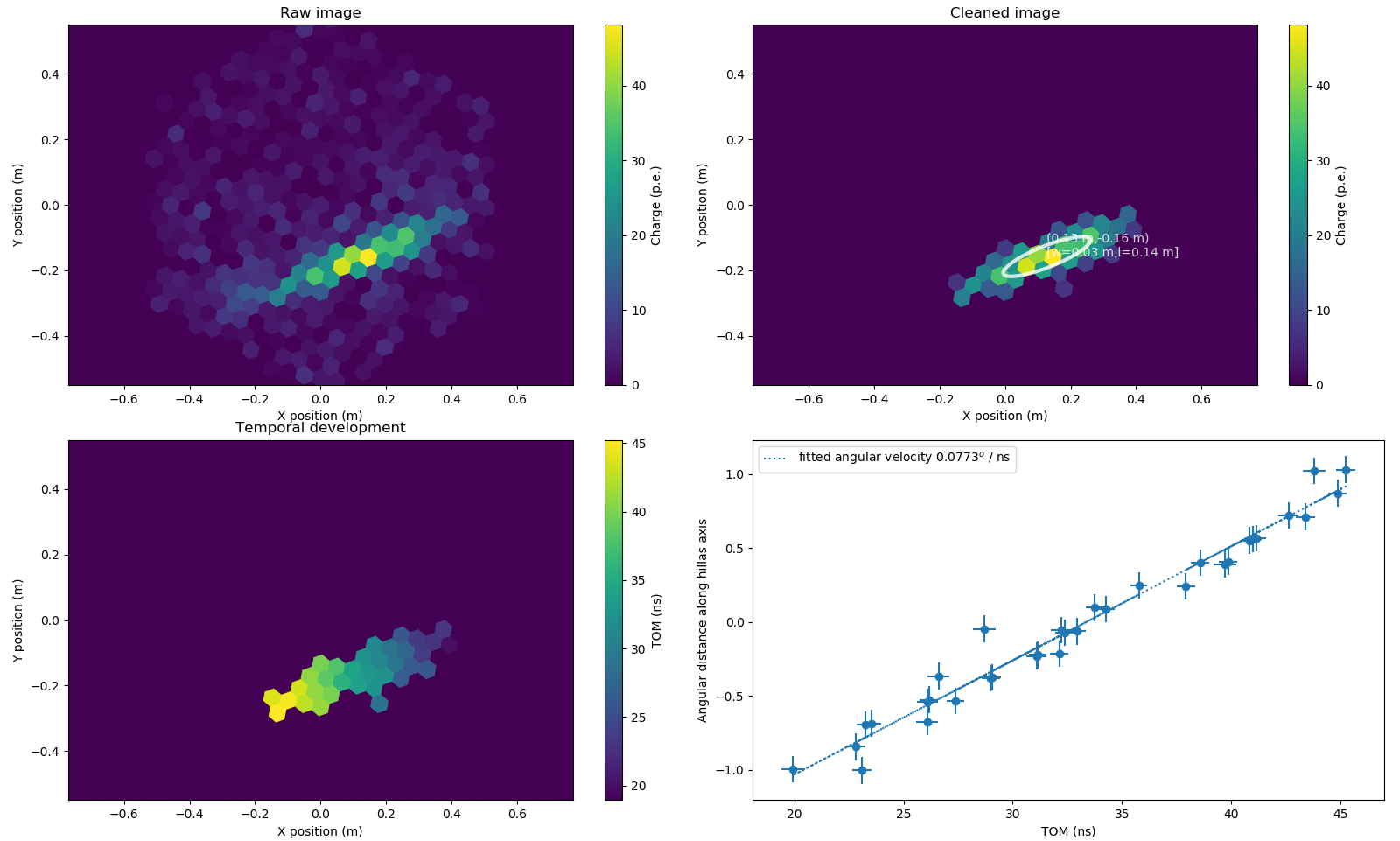}
\caption{Top left panel : photo-electron image of a $\gamma$-like event observed by the NectarCAM camera during Adlershof observation campaign. Top right panel : same event after the cleaning. The Hillas fit is superposed to the image. Bottom left panel : time of maximum for the cleaned pixels. Bottom right panel : the angular distance along the Hillas axis as a function of the time of maximum. The angular velocity of the shower image is fitted on the data.}
\label{shim}
\end{figure}

During the observation campaign in Adlershof, the NectarCAM camera successfully observed several hundred thousands  of atmospheric Cherenkov events. These data will provide precious information to understand the camera's performances. We will present here the steps of a simple atmospheric shower analysis, focusing on the value added by the timing information. Those steps are summarized in Fig.~\ref{shim}

During the observations, the TIB sends periodic trigger signals at a rate of 100 Hz. These events are used to estimate the baseline of the PMTs. The floating average of each 60 samples of the readout window for each pixels is subtracted to the signal from the real events. The  baseline subtraction is actually done during offline analysis but it is planned to be done during the writing of the data.

The next step of signal extraction is the estimation of the number of photo-electrons in each pixel. The ADC counts are summed on an sliding integration window of 16 ns aligned with the maximum of the waveform : 6 and 10 ns respectively before and after the maximum. The ADC sum is then divided by the gain of the PMT:
$$
Q = \sum_{i=-6}^{10} \left(\mathrm{ADC}_{i_\mathrm{max} + i} - P_{i_\mathrm{max} + i} \right) / G
$$
where Q is the charge i.e. the number of photo-electrons, P the baseline  and G is the gain.
Energetic showers may extend through several tens of nanosecond. The use of a sliding window allows to capture the full shower while using a relatively small integration window, thus maximizing the signal-to-noise ratio. 

Here, the cleaning of the image is done using a standard tailcut. This cleaning method uses a double threshold cut which takes into account the number of neighbouring pixels with a signal. To be kept in the image a pixel needs to be above 10 p.e. and have at least 3 neighbours above 5 p.e.. The timing could also be useful in the cleaning to exclude non coincident pixels. The two top panels of the figure \ref{shim} show a shower image before and after the cleaning.

Using the standard Hillas parametrization, we fit an ellipse on the shower image to get the axis of the shower development. The measurement of the time of maximum for the pixels in the shower image along the Hillas axis allows the reconstruction of the angular velocity of the shower projected in the focal plane. Bottom panels of Fig.\ref{shim} shows the temporal reconstruction of the shower, in the camera frame (bottom left) and along the Hillas axis (bottom right).

The temporal evolution of the shower and its influence on the reconstruction and the discrimination should be studied with Monte-Carlo simulation to fully understand the improvement induced by the use of the timing information.

\section{Conclusion}

We presented the calibration procedure for different parameters of the NectarCAM camera, focusing on the PMT's gain and the timing resolution. These calibration procedures and the results shown highlight the good performances of the camera. Despite the poor astronomical conditions in the vicinity of a big city like Berlin, the Adlershof campaign was a success and provided not only the first light of the NecatarCAM camera, but several hundred thousand of exploitable images from atmospheric Cherenkov events. The Calibration measurements done before the campaign and the one we are planning with the return of the camera in the dark room at CEA Paris-Saclay, will help us extracting the maximum amount of information from these valuable data.

\section*{ACKNOWLEDGMENTS}    
This work was conducted in the context of the CTA NectarCAM Project and has been carried out thanks to the support of the P2IO, OSUG2020 and OCEVU Labex (ANR-11-LABX-0060) as well as the A*MIDEX project (ANR-11-IDEX-0001-02) funded by the "Investissements d'Avenir" French government program managed by the ANR. We gratefully acknowledge support from the agencies and organisations listed in this page:\\ 
http://www.cta-observatory.org/consortium\_acknowledgments/


\end{document}